\documentclass[10pt,aip,sdy,superscriptaddress,floatfix,longbibliography,reprint]{revtex4-2}
\usepackage{graphicx}
\usepackage{dcolumn}
\usepackage{bm}
\usepackage{amsmath}
\usepackage{subfigure}
\usepackage{url} 
\usepackage{hyperref} 
\usepackage{xcolor}
\newcommand{\coso}{Cu$_2$OSeO$_3$}
\newcommand{\m}{\mathbf}

\renewcommand{\t}{\theta}

\newcommand{\be}{\begin{eqnarray}}
\newcommand{\ee}{\end{eqnarray}}

\begin{document}

\title{Helical spin dynamics in Cu$_2$OSeO$_3$ as measured with small-angle neutron scattering}

\author{Victor Ukleev}
\email{victor.ukleev@helmholtz-berlin.de}
\affiliation{Helmholtz-Zentrum Berlin f\"ur Materialien und Energie, D-12489 Berlin, Germany}
\author{Priya R. Baral}
\affiliation{Department of Applied Physics and Quantum-Phase Electronics Center, The University of Tokyo, Bunkyo, Tokyo 113-8656, Japan}
\author{Robert Cubitt}
\affiliation{Institut Laue–Langevin, 71 avenue des Martyrs, CS 20156, Grenoble, 38042 Cedex 9, France}
\author{Nina-Juliane Steinke}
\affiliation{Institut Laue–Langevin, 71 avenue des Martyrs, CS 20156, Grenoble, 38042 Cedex 9, France}
\author{Arnaud Magrez}
\affiliation{Institute of Physics, École Polytechnique Fédérale de Lausanne (EPFL), CH-1015 Lausanne, Switzerland}

\author{Oleg I. Utesov}
\affiliation{Center for Theoretical Physics of Complex Systems, Institute for Basic Science, Daejeon 34126, Republic of Korea}

\date{\today}

\begin{abstract}
The insulating chiral magnet Cu$_2$OSeO$_3$ exhibits a rich array of low-temperature magnetic phenomena, making it a prime candidate for the study of its spin dynamics. Using spin wave small-angle neutron scattering (SWSANS), we systematically investigated the temperature-dependent behavior of the helimagnon excitations in the field-polarized phase of \coso. Our measurements, spanning 5--55\,K, reveal the temperature evolution of spin-wave stiffness and damping constant with unprecedented resolution, facilitated by the insulating nature of \coso. These findings align with theoretical predictions and resolve discrepancies observed in previous studies, emphasizing the enhanced sensitivity of the SWSANS method. The results provide deeper insights into the fundamental magnetic properties of \coso, contributing to a broader understanding of chiral magnets.
\end{abstract}

\maketitle

\section{Introduction}

Recent studies have revealed a multitude of intriguing phenomena related to topological spin textures in cubic chiral magnets such as skyrmions \cite{bogdanov2005modulated,mühlbauer2009skyrmion,tokunaga2015new}, magnetic hedgehogs \cite{ kanazawa2017noncentrosymmetric,fujishiro2019topological}, merons \cite{yu2018transformation,yu2024spontaneous}, and hopfions \cite{zheng2023hopfion}. One of the prominent members of this family of materials is cubic chiral insulator and skyrmion host \coso, which with its unique spin-wave properties, offers a distinctive platform for applications in spintronic and magnonic devices \cite{stasinopoulos2017low,pollath2019ferromagnetic,aqeel2021microwave}. Notably, novel phases such as tilted conical spirals and disordered skyrmions have been recently identified in \coso~at low temperatures \cite{qian2018new,chacon2018observation,halder2018thermodynamic,bannenberg2019multiple,marchiori2024imaging,mehboodi2024observation}. Thus, our focus is on elucidating the temperature-dependent dynamics of helical spin excitations within the field-polarized regime of \coso. Refinement of its magnetic parameters such as spin-wave stiffness and damping can help fine-tune micromagnetic \cite{baral2022tuning} and \textit{ab-initio} \cite{janson2014quantum} theories of this remarkable material, which often suffer from a lack of reliable experimental input \cite{grytsiuk2019ab}.

In this study, we explore the spin-wave dynamics in Cu$_2$OSeO$_3$ using spin-wave small-angle neutron scattering (SWSANS) \cite{grigoriev2015spin}. High-resolution of SWSANS compared to classical inelastic neutron scattering experiments provide unique insights into low-energy spin dynamics in chiral helimagnets. Furthermore, the insulating nature of Cu$_2$OSeO$_3$ facilitates the extraction of damping constants, enabling a robust alignment with theoretical predictions. By applying an analytical framework, we offer quantitative assessments of both spin wave stiffness and damping constants across a wide temperature range from 5 to 55\,K ($T_\mathrm{C}=58$\,K \cite{seki2012observation}), comparing them with previous neutron scattering results.

\begin{figure*}
\caption{Magnetic field dependencies of SWSANS patterns at (a) 20\,K, (b) 38\,K and (c) 51\,K.}
\includegraphics[width=.95\linewidth]{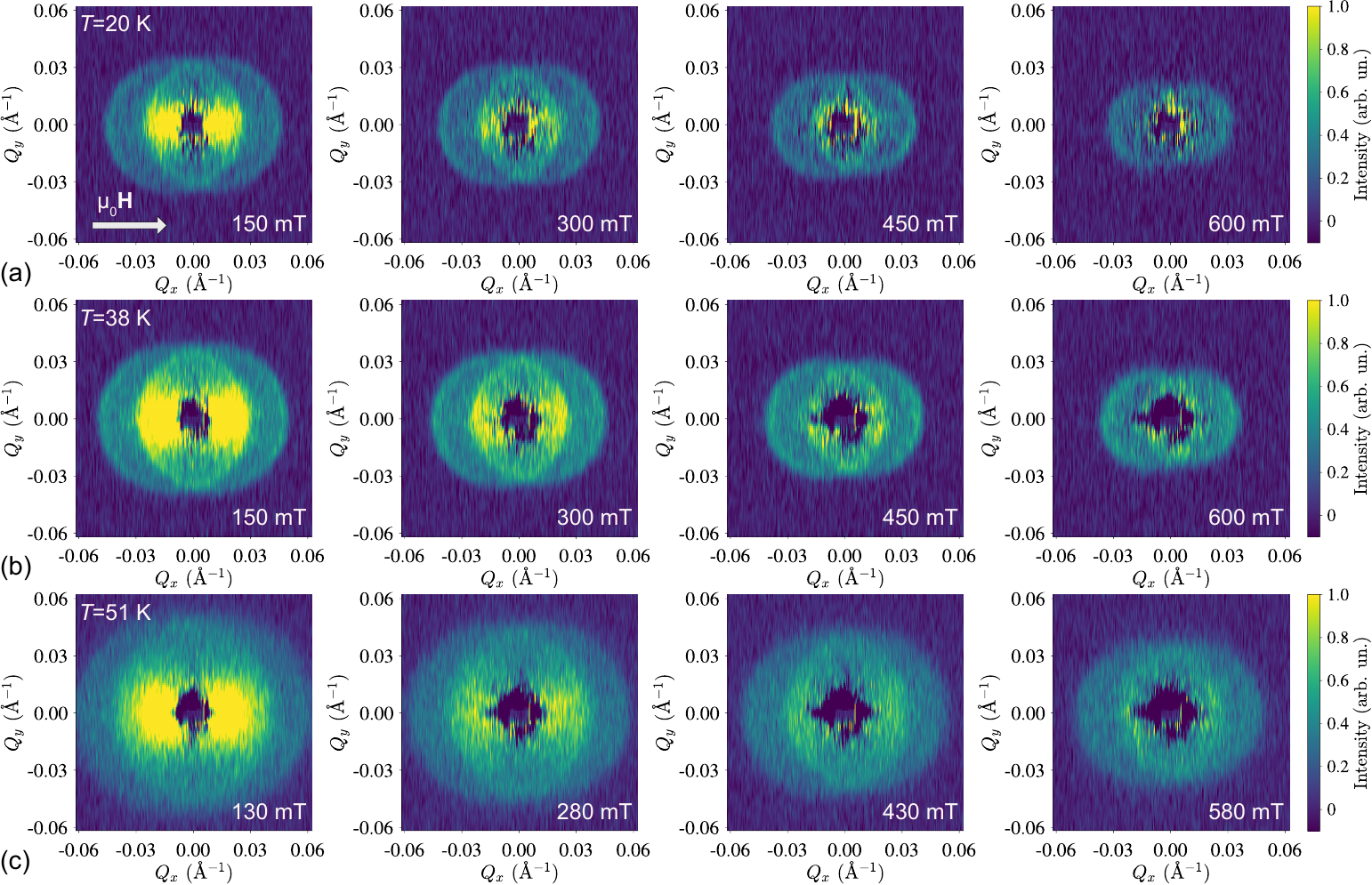}
\label{fig1}
\end{figure*}

\section{Experimental}

SWSANS measurements were carried out at D33 setup \cite{dewhurst2016small}, Institut Laue-Langevin (France) using neutrons with the wavelength $\lambda=5$\,\AA~and $\Delta \lambda/\lambda = 10\%$. The incoming beam was collimated over the distance of 7.8\,m and the detector was placed at the distance of 6.5\,m behind the sample. The acquisition time was 90 minutes per the SANS pattern. The magnetic field was controlled by the horizontal-field high-temperature superconducting magnet. Two-dimensional SANS data was reduced using GRASP software \cite{dewhurst2023graphical}.

 All SANS data presented in this article were obtained from $\sim$200\,mg disk-shaped single crystal containing (001) plane which has the diameter of 1\,cm and thickness of 2\,mm. The single crystal was grown by chemical vapor transport (CVT) reaction as described in Ref.~\cite{baral2022tuning}. The horizontal magnetic field was always aligned with the [110] crystal axis in the sample plane and perpendicular to the incoming neutron beam. Elastic SANS pattern measured in the helical magnetic phase at zero field in shown in the Appendix.

Measurements were performed for the temperature range 5--55\,K spanning through the entire phase diagram of \coso~including the A-phase \cite{seki2012observation} and novel low-temperature phases \cite{qian2018new,chacon2018observation}. At each temperature, SWSANS signal has been measured in 4 magnetic fields above the saturation ($H_{C2}$) in order to extract the linear dependence of the cut-off angle (see the next section for details). Background SANS signal was measured at 2\,K and 1\,T and subtracted from other data.

\section{Theoretical framework}

The spectrum of long-wavelength magnons in the fully polarized phase of cubic helimagnets reads~\cite{kataoka1987}
\begin{equation}   \label{spec1}
  \epsilon_\m{Q} = A_\mathrm{ex} (\m{Q} - \m{k}_\mathrm{s})^2 + g\mu_\mathrm{B} (H - H_\mathrm{C2}).
\end{equation}
Here $\m{Q}$ is the magnon wavevector, $A_\mathrm{ex}$ is the spin-wave stiffness, neglecting the anisotropic interactions $H_{C2} = A_\mathrm{ex} k^2_\mathrm{s}/g \mu_\mathrm{B}$ is the field of transition between the field-polarized and the conical phase, and $\m{k}_s$ corresponds to the spiral vector of the conical phase which is oriented along the field $\m{H}$. Noteworthy, cubic anisotropy and anisotropic exchange significantly influence \coso~ properties \cite{chacon2018observation,bannenberg2019multiple,baral202023direct}. However, in the fully polarized phase and our experimental geometry, their main effect is renormalization of $A_\mathrm{ex}$ and $H_\mathrm{C2}$ that preserves the form of Eq.~\eqref{spec1}. For definiteness, we choose the $x$-axis along $\m{H}$. For \coso, well below the ordering temperature $T_\mathrm{C} = 58$~K, $H_{C2} \sim 50$~mT \cite{seki2012observation} and $k_s =0.0102$~nm$^{-1}$ which corresponds to the spiral pitch \emph{ca.} $62$~nm \cite{adams2012long}.

The crucial parameter for SWSANS relates the energy of the incident neutrons with the magnetic system energy scale~\cite{grigoriev2015spin}. Explicitly, we define $\t_0 = E_i / A_\mathrm{ex} k^2_i $ where $E_i = \hbar^2 k^2_i/2m_n$. Then, for the spectrum~\eqref{spec1}, the kinematic conditions for the scattering are fulfilled only if the scattering angle in the detector plane $\t=(\t_x,\t_y)$ satisfies $\t^2_\mathrm{rel} \equiv (\t_x - \t_B)^2 +\t^2_y \leq \t^2_C$ where the Bragg angle $\t_B = k_s/k_i$ and the \emph{cut-off angle} is given by~\cite{grigoriev2015spin} 
\be \label{cutoff1}
  \t^2_C = \t^2_0 - \frac{g \mu_B (H-H_{C2}) \t_0}{E_i}.
\ee
By performing a linear fit to the field dependence of the cut-off angle, crucial parameter $\theta_0$ can be readily extracted and, consequently, $A_{\mathrm ex}$ value can be determined.

\begin{figure*}
\caption{Radially averaged profiles of SWSANS intensity measured at (a) 20\,K, (b) 38\,K and (c) 51\,K and at different magnetic fields. Solid lines correspond to the fit according to Eq.~\eqref{sigma1}.}
\includegraphics[width=.95\linewidth]{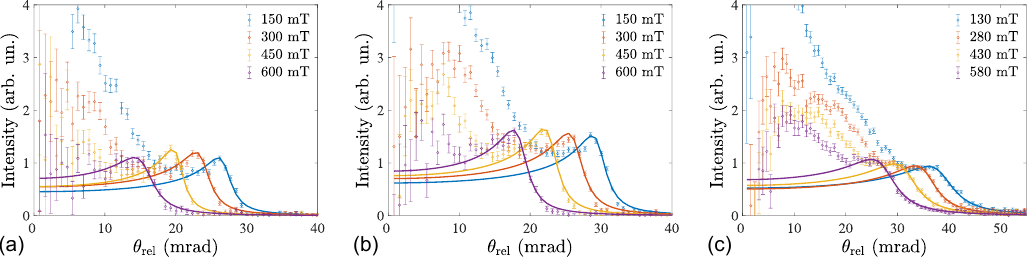}
\label{fig2}
\end{figure*}

Introducing the magnon spectral line broadening $\Gamma$ allows relaxing the energy conservation law. The physical mechanisms behind this quantity in dielectrics can be magnon-magnon scattering, magnetoelastic coupling, and lattice impurities. In Ref.~\cite{ukleev2022spin} it was shown that in order to account for the magnon damping, one can use the formula
\be \label{sigma1}
  \sigma(\t) \propto \textrm{Im} \frac{1}{\sqrt{\t^2_\mathrm{rel} -\t^2_C - i \gamma}}
\ee
for the SWSANS data interpretation. Here $\gamma =\t_0 \Gamma/E_i$ is the dimensionless broadening for magnons at the cut-off which have energy $\epsilon = 2 E_i \t_0$ (in practice, $\Gamma$ also includes a contribution from the device resolution).

\section{Results and Discussion}

SWSANS patterns from the Cu$_2$OSeO$_3$ crystal were measured at several temperatures. The intensity of the inelastic scattering is significantly increased compared to the previous study by Grigoriev, \textit{et al.} \cite{grigoriev2019spin} thanks to the larger sample volume.
Typical SWSANS data measured at 20\,K, 38\,K and 51\,K at different magnetic field magnitudes are shown in Figure \ref{fig1}. Despite the development of a strong exchange anisotropy at low temperatures in Cu$_2$OSeO$_3$ \cite{baral202023direct}, its influence is negligible for the present experimental geometry, and its influence on SWSANS will be a subject of our future works. At each temperature, the onset of the shrinking SWSANS circles allows to extract the exchange stiffness parameter $A_\mathrm{ex}$. From the two-dimensional SANS images, one can already note that the scattering intensity is somewhat enhanced on the edge of the circle, following the theoretical prediction made in Ref. \cite{ukleev2022spin}. 

For quantification of the $A_\mathrm{ex}$, SWSANS intensity profiles were extracted for each temperature and field point measured. Typical radially averaged profiles of SWSANS intensity are shown in Figure \ref{fig2}. Note, that the radial averaging was done from the center of the elastic helical Bragg peaks at the angle $\theta_\mathrm{B}$. SWSANS radial profiles were fitted to the model according to Eq.\eqref{sigma1} (solid lines in Fig. \ref{fig2}). A peak of SWSANS intensity is clearly observed just before the cutoff angle at lower temperatures, which perfectly agrees with the model. The peak has been predicted in Ref. \cite{ukleev2022spin}, however, could not be detected experimentally in earlier SWSANS works due to the strong SW damping in metallic chiral magnets MnSi \cite{grigoriev2015spin}, Mn$_{1-x}$Fe$_x$Si \cite{grigoriev2018spin},FeGe \cite{siegfried2017spin}, Fe$_{1-x}$Co$_x$Si \cite{grigoriev2019spin2} and Co$_8$Zn$_8$Mn$_4$ \cite{ukleev2022spin}. In the previous study on \coso, the bump was not clearly discernible due to the smaller sample size and was also not accounted for by the theory proposed in Ref.~\cite{grigoriev2019spin}. However, we would like to point out that it was experimentally observed using polarized neutron scattering on spin waves in amorphous ferromagnets \cite{okorokov1986study,deriglazov1992study}.

\begin{figure}
\caption{(a) Exchange stiffness and (b) damping parameters of \coso~obtained from the SWSANS data and modeling. Errorbars in the panel (a) are smaller than the symbol size for temperatures below 50\,K}.
\includegraphics[width=.9\linewidth]{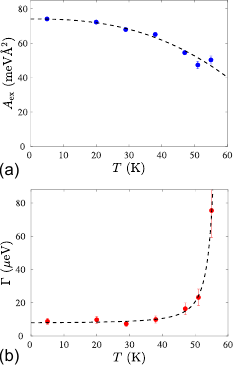}
\label{fig3}
\end{figure}

Exchange stiffness and spin-wave damping parameters obtained from the fit are shown in Fig.~\ref{fig3}(a) and (b), respectively. The dashed line in Fig.~\ref{fig3}(a) represents the fit according to the law
\be
  A = A_0 \left[ 1 - c \left( \frac{T}{T_\mathrm{C}} \right)^{5/2} \right].
\ee
The resultant fit matches the experimental data well. Here the parameter $c$ is a constant of an order of 1 which can be obtained by integrating the magnon interaction and their Bose-Einstein occupation numbers over the entire Brillouin zone, and it cannot be determined without a precise microscopic model. The fitted parameters, $A_0 = 74.1\pm0.3$\,meV\AA$^2$ and $c = 0.42\pm0.05$, align very well with those from the previous SWSANS study reported by \cite{grigoriev2019spin}. Exchange stiffness parameter in previous cold neutron inelastic scattering experiments was extracted from the parabolic fit of the dispersion curves \cite{luo2020low,portnichenko2016magnon}. Interestingly, a similar stiffness to our value was found in Refs. \cite{tucker2016spin,luo2020low}, while somewhat lower values of 50 meV\AA$^2$~ were also reported \cite{portnichenko2016magnon}. We believe that the ultimate resolution of our SWSANS probe resolves the discrepancy in favor of the higher value. Regarding the damping, Fig. \ref{fig3}(b) shows that it naturally increases with temperature, becoming significantly large near $T_\mathrm{C}$, where magnons cease to be well-defined quasiparticles (in particular, power-law fit reveals $\Gamma \propto (T_\mathrm{C} - T)^{-2}$ dependence). The broadening of the cutoff angle due to the finite experimental resolution can be estimated as $\gamma_r=\theta_\mathrm{C}\delta \theta$, given the resolution of 4\,mrad mainly due to the 10\% neutron wavelength bandwidth. Hence the corresponding minimal measurable damping parameter $\Gamma_{\mathrm{min}}\approx 8$\,$\mu$eV which which explains well the saturation observed below 40\,K (Fig. \ref{fig3}b).

\section{Conclusion}

We have presented a detailed study of the spin-wave dynamics in Cu$_2$OSeO$_3$ using small-angle neutron scattering, focusing on the temperature dependence of spin-wave stiffness and damping constants. Our findings demonstrate a significant enhancement in the sensitivity of SANS due to the larger sample size and the insulating nature of the material, enabling the detection of subtle features predicted theoretically but previously unresolved. The extracted exchange stiffness values show excellent agreement with established models, while the damping parameters highlight the low intrinsic spin-wave damping in Cu$_2$OSeO$_3$ compared to metallic chiral magnets. These results not only provide a comprehensive understanding of the helimagnon dynamics in Cu$_2$OSeO$_3$, but also establish the potential of advanced SANS techniques for probing spin-wave phenomena in other complex magnetic systems.

\section*{Acknowledgements}

Authors thank S. V. Grigoriev, K. A. Pschenichnyi, and J. S. White for helpful discussions and M. Bonnaud for the technical support. We acknowledge Institut Laue-Langevin (France) for the SANS beamtime provided according to the proposal 5-41-1226 \cite{illdata}. P.R.B. acknowledges Swiss National Science Foundation (SNSF) Postdoc.Mobility grant P500PT\_217697 for financial assistance. O.I.U. acknowledges financial support from the Institute for Basic Science (IBS) in the Republic of Korea through Projects No. IBS-R024-D1 and No. IBS-R024-Y3.

\section*{Author Declarations}
\subsection*{Conflict of Interest}
The authors have no conflicts to disclose.
\subsection*{Author Contributions}
V.U., P.R.B., R.C., N.-J. S. performed SANS experiments, P.R.B. and A.M. synthesized the sample, V.U. and O.I.U. analyzed the data, O.I.U. wrote the theory, V.U., P.R.B, O.I.U., jointly conceived the project.
\section*{Data availability Statement}

Data is available from V.U. upon request and from the ILL data repository \cite{illdata} after the embargo period.  

\section*{Appendix}

\begin{figure}
\caption{Elastic SANS pattern measured at zero magnetic field at $T=2$\,K in the helical phase.}
\includegraphics[width=.9\linewidth]{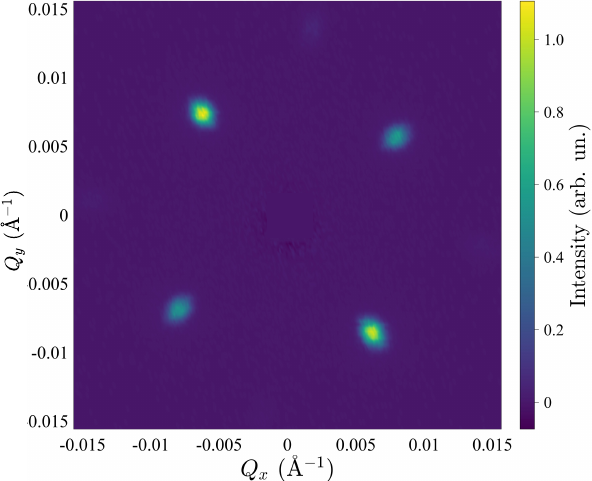}
\label{fig4}
\end{figure}

Elastic SANS pattern measured at zero magnetic field at the base temperature $T=2$\,K is shown in Fig. \ref{fig4}. The pattern manifests a two-domain helical state with the propagation vectors along [100] and [010] crystal axes, respectively.

\bibliography{biblio}

\begin{thebibliography}{38}%
\makeatletter
\providecommand \@ifxundefined [1]{%
 \@ifx{#1\undefined}
}%
\providecommand \@ifnum [1]{%
 \ifnum #1\expandafter \@firstoftwo
 \else \expandafter \@secondoftwo
 \fi
}%
\providecommand \@ifx [1]{%
 \ifx #1\expandafter \@firstoftwo
 \else \expandafter \@secondoftwo
 \fi
}%
\providecommand \natexlab [1]{#1}%
\providecommand \enquote  [1]{``#1''}%
\providecommand \bibnamefont  [1]{#1}%
\providecommand \bibfnamefont [1]{#1}%
\providecommand \citenamefont [1]{#1}%
\providecommand \href@noop [0]{\@secondoftwo}%
\providecommand \href [0]{\begingroup \@sanitize@url \@href}%
\providecommand \@href[1]{\@@startlink{#1}\@@href}%
\providecommand \@@href[1]{\endgroup#1\@@endlink}%
\providecommand \@sanitize@url [0]{\catcode `\\12\catcode `\$12\catcode `\&12\catcode `\#12\catcode `\^12\catcode `\_12\catcode `\%12\relax}%
\providecommand \@@startlink[1]{}%
\providecommand \@@endlink[0]{}%
\providecommand \url  [0]{\begingroup\@sanitize@url \@url }%
\providecommand \@url [1]{\endgroup\@href {#1}{\urlprefix }}%
\providecommand \urlprefix  [0]{URL }%
\providecommand \Eprint [0]{\href }%
\providecommand \doibase [0]{https://doi.org/}%
\providecommand \selectlanguage [0]{\@gobble}%
\providecommand \bibinfo  [0]{\@secondoftwo}%
\providecommand \bibfield  [0]{\@secondoftwo}%
\providecommand \translation [1]{[#1]}%
\providecommand \BibitemOpen [0]{}%
\providecommand \bibitemStop [0]{}%
\providecommand \bibitemNoStop [0]{.\EOS\space}%
\providecommand \EOS [0]{\spacefactor3000\relax}%
\providecommand \BibitemShut  [1]{\csname bibitem#1\endcsname}%
\let\auto@bib@innerbib\@empty
\bibitem [{\citenamefont {Bogdanov}, \citenamefont {R{\"o}ssler},\ and\ \citenamefont {Pfleiderer}(2005)}]{bogdanov2005modulated}%
  \BibitemOpen
  \bibfield  {author} {\bibinfo {author} {\bibfnamefont {A.}~\bibnamefont {Bogdanov}}, \bibinfo {author} {\bibfnamefont {U.}~\bibnamefont {R{\"o}ssler}},\ and\ \bibinfo {author} {\bibfnamefont {C.}~\bibnamefont {Pfleiderer}},\ }\bibfield  {title} {\enquote {\bibinfo {title} {Modulated and localized structures in cubic helimagnets},}\ }\href@noop {} {\bibfield  {journal} {\bibinfo  {journal} {Physica B: Condensed Matter}\ }\textbf {\bibinfo {volume} {359}},\ \bibinfo {pages} {1162--1164} (\bibinfo {year} {2005})}\BibitemShut {NoStop}%
\bibitem [{\citenamefont {Mühlbauer}\ \emph {et~al.}(2009)\citenamefont {Mühlbauer}, \citenamefont {Binz}, \citenamefont {Jonietz}, \citenamefont {Pfleiderer}, \citenamefont {Rosch}, \citenamefont {Neubauer}, \citenamefont {Georgii},\ and\ \citenamefont {Böni}}]{mühlbauer2009skyrmion}%
  \BibitemOpen
  \bibfield  {author} {\bibinfo {author} {\bibfnamefont {S.}~\bibnamefont {Mühlbauer}}, \bibinfo {author} {\bibfnamefont {B.}~\bibnamefont {Binz}}, \bibinfo {author} {\bibfnamefont {F.}~\bibnamefont {Jonietz}}, \bibinfo {author} {\bibfnamefont {C.}~\bibnamefont {Pfleiderer}}, \bibinfo {author} {\bibfnamefont {A.}~\bibnamefont {Rosch}}, \bibinfo {author} {\bibfnamefont {A.}~\bibnamefont {Neubauer}}, \bibinfo {author} {\bibfnamefont {R.}~\bibnamefont {Georgii}},\ and\ \bibinfo {author} {\bibfnamefont {P.}~\bibnamefont {Böni}},\ }\bibfield  {title} {\enquote {\bibinfo {title} {Skyrmion lattice in a chiral magnet},}\ }\href@noop {} {\bibfield  {journal} {\bibinfo  {journal} {Science}\ }\textbf {\bibinfo {volume} {323}},\ \bibinfo {pages} {915--919} (\bibinfo {year} {2009})}\BibitemShut {NoStop}%
\bibitem [{\citenamefont {Tokunaga}\ \emph {et~al.}(2015)\citenamefont {Tokunaga}, \citenamefont {Yu}, \citenamefont {White}, \citenamefont {R{\o}nnow}, \citenamefont {Morikawa}, \citenamefont {Taguchi},\ and\ \citenamefont {Tokura}}]{tokunaga2015new}%
  \BibitemOpen
  \bibfield  {author} {\bibinfo {author} {\bibfnamefont {Y.}~\bibnamefont {Tokunaga}}, \bibinfo {author} {\bibfnamefont {X.}~\bibnamefont {Yu}}, \bibinfo {author} {\bibfnamefont {J.}~\bibnamefont {White}}, \bibinfo {author} {\bibfnamefont {H.~M.}\ \bibnamefont {R{\o}nnow}}, \bibinfo {author} {\bibfnamefont {D.}~\bibnamefont {Morikawa}}, \bibinfo {author} {\bibfnamefont {Y.}~\bibnamefont {Taguchi}},\ and\ \bibinfo {author} {\bibfnamefont {Y.}~\bibnamefont {Tokura}},\ }\bibfield  {title} {\enquote {\bibinfo {title} {A new class of chiral materials hosting magnetic skyrmions beyond room temperature},}\ }\href@noop {} {\bibfield  {journal} {\bibinfo  {journal} {Nature Communications}\ }\textbf {\bibinfo {volume} {6}},\ \bibinfo {pages} {7638} (\bibinfo {year} {2015})}\BibitemShut {NoStop}%
\bibitem [{\citenamefont {Kanazawa}, \citenamefont {Seki},\ and\ \citenamefont {Tokura}(2017)}]{kanazawa2017noncentrosymmetric}%
  \BibitemOpen
  \bibfield  {author} {\bibinfo {author} {\bibfnamefont {N.}~\bibnamefont {Kanazawa}}, \bibinfo {author} {\bibfnamefont {S.}~\bibnamefont {Seki}},\ and\ \bibinfo {author} {\bibfnamefont {Y.}~\bibnamefont {Tokura}},\ }\bibfield  {title} {\enquote {\bibinfo {title} {Noncentrosymmetric magnets hosting magnetic skyrmions},}\ }\href@noop {} {\bibfield  {journal} {\bibinfo  {journal} {Advanced Materials}\ }\textbf {\bibinfo {volume} {29}},\ \bibinfo {pages} {1603227} (\bibinfo {year} {2017})}\BibitemShut {NoStop}%
\bibitem [{\citenamefont {Fujishiro}\ \emph {et~al.}(2019)\citenamefont {Fujishiro}, \citenamefont {Kanazawa}, \citenamefont {Nakajima}, \citenamefont {Yu}, \citenamefont {Ohishi}, \citenamefont {Kawamura}, \citenamefont {Kakurai}, \citenamefont {Arima}, \citenamefont {Mitamura}, \citenamefont {Miyake} \emph {et~al.}}]{fujishiro2019topological}%
  \BibitemOpen
  \bibfield  {author} {\bibinfo {author} {\bibfnamefont {Y.}~\bibnamefont {Fujishiro}}, \bibinfo {author} {\bibfnamefont {N.}~\bibnamefont {Kanazawa}}, \bibinfo {author} {\bibfnamefont {T.}~\bibnamefont {Nakajima}}, \bibinfo {author} {\bibfnamefont {X.}~\bibnamefont {Yu}}, \bibinfo {author} {\bibfnamefont {K.}~\bibnamefont {Ohishi}}, \bibinfo {author} {\bibfnamefont {Y.}~\bibnamefont {Kawamura}}, \bibinfo {author} {\bibfnamefont {K.}~\bibnamefont {Kakurai}}, \bibinfo {author} {\bibfnamefont {T.}~\bibnamefont {Arima}}, \bibinfo {author} {\bibfnamefont {H.}~\bibnamefont {Mitamura}}, \bibinfo {author} {\bibfnamefont {A.}~\bibnamefont {Miyake}}, \emph {et~al.},\ }\bibfield  {title} {\enquote {\bibinfo {title} {Topological transitions among skyrmion-and hedgehog-lattice states in cubic chiral magnets},}\ }\href@noop {} {\bibfield  {journal} {\bibinfo  {journal} {Nature Communications}\ }\textbf {\bibinfo {volume} {10}},\ \bibinfo {pages} {1059} (\bibinfo {year} {2019})}\BibitemShut {NoStop}%
\bibitem [{\citenamefont {Yu}\ \emph {et~al.}(2018)\citenamefont {Yu}, \citenamefont {Koshibae}, \citenamefont {Tokunaga}, \citenamefont {Shibata}, \citenamefont {Taguchi}, \citenamefont {Nagaosa},\ and\ \citenamefont {Tokura}}]{yu2018transformation}%
  \BibitemOpen
  \bibfield  {author} {\bibinfo {author} {\bibfnamefont {X.}~\bibnamefont {Yu}}, \bibinfo {author} {\bibfnamefont {W.}~\bibnamefont {Koshibae}}, \bibinfo {author} {\bibfnamefont {Y.}~\bibnamefont {Tokunaga}}, \bibinfo {author} {\bibfnamefont {K.}~\bibnamefont {Shibata}}, \bibinfo {author} {\bibfnamefont {Y.}~\bibnamefont {Taguchi}}, \bibinfo {author} {\bibfnamefont {N.}~\bibnamefont {Nagaosa}},\ and\ \bibinfo {author} {\bibfnamefont {Y.}~\bibnamefont {Tokura}},\ }\bibfield  {title} {\enquote {\bibinfo {title} {Transformation between meron and skyrmion topological spin textures in a chiral magnet},}\ }\href@noop {} {\bibfield  {journal} {\bibinfo  {journal} {Nature}\ }\textbf {\bibinfo {volume} {564}},\ \bibinfo {pages} {95--98} (\bibinfo {year} {2018})}\BibitemShut {NoStop}%
\bibitem [{\citenamefont {Yu}\ \emph {et~al.}(2024)\citenamefont {Yu}, \citenamefont {Kanazawa}, \citenamefont {Zhang}, \citenamefont {Takahashi}, \citenamefont {Iakoubovskii}, \citenamefont {Nakajima}, \citenamefont {Tanigaki}, \citenamefont {Mochizuki},\ and\ \citenamefont {Tokura}}]{yu2024spontaneous}%
  \BibitemOpen
  \bibfield  {author} {\bibinfo {author} {\bibfnamefont {X.}~\bibnamefont {Yu}}, \bibinfo {author} {\bibfnamefont {N.}~\bibnamefont {Kanazawa}}, \bibinfo {author} {\bibfnamefont {X.}~\bibnamefont {Zhang}}, \bibinfo {author} {\bibfnamefont {Y.}~\bibnamefont {Takahashi}}, \bibinfo {author} {\bibfnamefont {K.~V.}\ \bibnamefont {Iakoubovskii}}, \bibinfo {author} {\bibfnamefont {K.}~\bibnamefont {Nakajima}}, \bibinfo {author} {\bibfnamefont {T.}~\bibnamefont {Tanigaki}}, \bibinfo {author} {\bibfnamefont {M.}~\bibnamefont {Mochizuki}},\ and\ \bibinfo {author} {\bibfnamefont {Y.}~\bibnamefont {Tokura}},\ }\bibfield  {title} {\enquote {\bibinfo {title} {Spontaneous vortex-antivortex pairs and their topological transitions in a chiral-lattice magnet},}\ }\href@noop {} {\bibfield  {journal} {\bibinfo  {journal} {Advanced Materials}\ }\textbf {\bibinfo {volume} {36}},\ \bibinfo {pages} {2306441} (\bibinfo {year} {2024})}\BibitemShut {NoStop}%
\bibitem [{\citenamefont {Zheng}\ \emph {et~al.}(2023)\citenamefont {Zheng}, \citenamefont {Kiselev}, \citenamefont {Rybakov}, \citenamefont {Yang}, \citenamefont {Shi}, \citenamefont {Bl{\"u}gel},\ and\ \citenamefont {Dunin-Borkowski}}]{zheng2023hopfion}%
  \BibitemOpen
  \bibfield  {author} {\bibinfo {author} {\bibfnamefont {F.}~\bibnamefont {Zheng}}, \bibinfo {author} {\bibfnamefont {N.~S.}\ \bibnamefont {Kiselev}}, \bibinfo {author} {\bibfnamefont {F.~N.}\ \bibnamefont {Rybakov}}, \bibinfo {author} {\bibfnamefont {L.}~\bibnamefont {Yang}}, \bibinfo {author} {\bibfnamefont {W.}~\bibnamefont {Shi}}, \bibinfo {author} {\bibfnamefont {S.}~\bibnamefont {Bl{\"u}gel}},\ and\ \bibinfo {author} {\bibfnamefont {R.~E.}\ \bibnamefont {Dunin-Borkowski}},\ }\bibfield  {title} {\enquote {\bibinfo {title} {Hopfion rings in a cubic chiral magnet},}\ }\href@noop {} {\bibfield  {journal} {\bibinfo  {journal} {Nature}\ }\textbf {\bibinfo {volume} {623}},\ \bibinfo {pages} {718--723} (\bibinfo {year} {2023})}\BibitemShut {NoStop}%
\bibitem [{\citenamefont {Stasinopoulos}\ \emph {et~al.}(2017)\citenamefont {Stasinopoulos}, \citenamefont {Weichselbaumer}, \citenamefont {Bauer}, \citenamefont {Waizner}, \citenamefont {Berger}, \citenamefont {Maendl}, \citenamefont {Garst}, \citenamefont {Pfleiderer},\ and\ \citenamefont {Grundler}}]{stasinopoulos2017low}%
  \BibitemOpen
  \bibfield  {author} {\bibinfo {author} {\bibfnamefont {I.}~\bibnamefont {Stasinopoulos}}, \bibinfo {author} {\bibfnamefont {S.}~\bibnamefont {Weichselbaumer}}, \bibinfo {author} {\bibfnamefont {A.}~\bibnamefont {Bauer}}, \bibinfo {author} {\bibfnamefont {J.}~\bibnamefont {Waizner}}, \bibinfo {author} {\bibfnamefont {H.}~\bibnamefont {Berger}}, \bibinfo {author} {\bibfnamefont {S.}~\bibnamefont {Maendl}}, \bibinfo {author} {\bibfnamefont {M.}~\bibnamefont {Garst}}, \bibinfo {author} {\bibfnamefont {C.}~\bibnamefont {Pfleiderer}},\ and\ \bibinfo {author} {\bibfnamefont {D.}~\bibnamefont {Grundler}},\ }\bibfield  {title} {\enquote {\bibinfo {title} {Low spin wave damping in the insulating chiral magnet {Cu$_2$OSeO$_3$}},}\ }\href@noop {} {\bibfield  {journal} {\bibinfo  {journal} {Applied Physics Letters}\ }\textbf {\bibinfo {volume} {111}} (\bibinfo {year} {2017})}\BibitemShut {NoStop}%
\bibitem [{\citenamefont {P{\"o}llath}\ \emph {et~al.}(2019)\citenamefont {P{\"o}llath}, \citenamefont {Aqeel}, \citenamefont {Bauer}, \citenamefont {Luo}, \citenamefont {Ryll}, \citenamefont {Radu}, \citenamefont {Pfleiderer}, \citenamefont {Woltersdorf},\ and\ \citenamefont {Back}}]{pollath2019ferromagnetic}%
  \BibitemOpen
  \bibfield  {author} {\bibinfo {author} {\bibfnamefont {S.}~\bibnamefont {P{\"o}llath}}, \bibinfo {author} {\bibfnamefont {A.}~\bibnamefont {Aqeel}}, \bibinfo {author} {\bibfnamefont {A.}~\bibnamefont {Bauer}}, \bibinfo {author} {\bibfnamefont {C.}~\bibnamefont {Luo}}, \bibinfo {author} {\bibfnamefont {H.}~\bibnamefont {Ryll}}, \bibinfo {author} {\bibfnamefont {F.}~\bibnamefont {Radu}}, \bibinfo {author} {\bibfnamefont {C.}~\bibnamefont {Pfleiderer}}, \bibinfo {author} {\bibfnamefont {G.}~\bibnamefont {Woltersdorf}},\ and\ \bibinfo {author} {\bibfnamefont {C.~H.}\ \bibnamefont {Back}},\ }\bibfield  {title} {\enquote {\bibinfo {title} {Ferromagnetic resonance with magnetic phase selectivity by means of resonant elastic x-ray scattering on a chiral magnet},}\ }\href@noop {} {\bibfield  {journal} {\bibinfo  {journal} {Physical Review Letters}\ }\textbf {\bibinfo {volume} {123}},\ \bibinfo {pages} {167201} (\bibinfo {year} {2019})}\BibitemShut {NoStop}%
\bibitem [{\citenamefont {Aqeel}\ \emph {et~al.}(2021)\citenamefont {Aqeel}, \citenamefont {Sahliger}, \citenamefont {Taniguchi}, \citenamefont {M{\"a}ndl}, \citenamefont {Mettus}, \citenamefont {Berger}, \citenamefont {Bauer}, \citenamefont {Garst}, \citenamefont {Pfleiderer},\ and\ \citenamefont {Back}}]{aqeel2021microwave}%
  \BibitemOpen
  \bibfield  {author} {\bibinfo {author} {\bibfnamefont {A.}~\bibnamefont {Aqeel}}, \bibinfo {author} {\bibfnamefont {J.}~\bibnamefont {Sahliger}}, \bibinfo {author} {\bibfnamefont {T.}~\bibnamefont {Taniguchi}}, \bibinfo {author} {\bibfnamefont {S.}~\bibnamefont {M{\"a}ndl}}, \bibinfo {author} {\bibfnamefont {D.}~\bibnamefont {Mettus}}, \bibinfo {author} {\bibfnamefont {H.}~\bibnamefont {Berger}}, \bibinfo {author} {\bibfnamefont {A.}~\bibnamefont {Bauer}}, \bibinfo {author} {\bibfnamefont {M.}~\bibnamefont {Garst}}, \bibinfo {author} {\bibfnamefont {C.}~\bibnamefont {Pfleiderer}},\ and\ \bibinfo {author} {\bibfnamefont {C.~H.}\ \bibnamefont {Back}},\ }\bibfield  {title} {\enquote {\bibinfo {title} {Microwave spectroscopy of the low-temperature skyrmion state in {Cu$_2$OSeO$_3$}},}\ }\href@noop {} {\bibfield  {journal} {\bibinfo  {journal} {Physical Review Letters}\ }\textbf {\bibinfo {volume} {126}},\ \bibinfo {pages} {017202} (\bibinfo {year} {2021})}\BibitemShut {NoStop}%
\bibitem [{\citenamefont {Qian}\ \emph {et~al.}(2018)\citenamefont {Qian}, \citenamefont {Bannenberg}, \citenamefont {Wilhelm}, \citenamefont {Chaboussant}, \citenamefont {Debeer-Schmitt}, \citenamefont {Schmidt}, \citenamefont {Aqeel}, \citenamefont {Palstra}, \citenamefont {Br{\"u}ck}, \citenamefont {Lefering} \emph {et~al.}}]{qian2018new}%
  \BibitemOpen
  \bibfield  {author} {\bibinfo {author} {\bibfnamefont {F.}~\bibnamefont {Qian}}, \bibinfo {author} {\bibfnamefont {L.~J.}\ \bibnamefont {Bannenberg}}, \bibinfo {author} {\bibfnamefont {H.}~\bibnamefont {Wilhelm}}, \bibinfo {author} {\bibfnamefont {G.}~\bibnamefont {Chaboussant}}, \bibinfo {author} {\bibfnamefont {L.~M.}\ \bibnamefont {Debeer-Schmitt}}, \bibinfo {author} {\bibfnamefont {M.~P.}\ \bibnamefont {Schmidt}}, \bibinfo {author} {\bibfnamefont {A.}~\bibnamefont {Aqeel}}, \bibinfo {author} {\bibfnamefont {T.~T.}\ \bibnamefont {Palstra}}, \bibinfo {author} {\bibfnamefont {E.}~\bibnamefont {Br{\"u}ck}}, \bibinfo {author} {\bibfnamefont {A.~J.}\ \bibnamefont {Lefering}}, \emph {et~al.},\ }\bibfield  {title} {\enquote {\bibinfo {title} {New magnetic phase of the chiral skyrmion material {Cu$_2$OSeO$_3$}},}\ }\href@noop {} {\bibfield  {journal} {\bibinfo  {journal} {Science Advances}\ }\textbf {\bibinfo {volume} {4}},\ \bibinfo {pages} {eaat7323} (\bibinfo {year} {2018})}\BibitemShut {NoStop}%
\bibitem [{\citenamefont {Chacon}\ \emph {et~al.}(2018)\citenamefont {Chacon}, \citenamefont {Heinen}, \citenamefont {Halder}, \citenamefont {Bauer}, \citenamefont {Simeth}, \citenamefont {M{\"u}hlbauer}, \citenamefont {Berger}, \citenamefont {Garst}, \citenamefont {Rosch},\ and\ \citenamefont {Pfleiderer}}]{chacon2018observation}%
  \BibitemOpen
  \bibfield  {author} {\bibinfo {author} {\bibfnamefont {A.}~\bibnamefont {Chacon}}, \bibinfo {author} {\bibfnamefont {L.}~\bibnamefont {Heinen}}, \bibinfo {author} {\bibfnamefont {M.}~\bibnamefont {Halder}}, \bibinfo {author} {\bibfnamefont {A.}~\bibnamefont {Bauer}}, \bibinfo {author} {\bibfnamefont {W.}~\bibnamefont {Simeth}}, \bibinfo {author} {\bibfnamefont {S.}~\bibnamefont {M{\"u}hlbauer}}, \bibinfo {author} {\bibfnamefont {H.}~\bibnamefont {Berger}}, \bibinfo {author} {\bibfnamefont {M.}~\bibnamefont {Garst}}, \bibinfo {author} {\bibfnamefont {A.}~\bibnamefont {Rosch}},\ and\ \bibinfo {author} {\bibfnamefont {C.}~\bibnamefont {Pfleiderer}},\ }\bibfield  {title} {\enquote {\bibinfo {title} {Observation of two independent skyrmion phases in a chiral magnetic material},}\ }\href@noop {} {\bibfield  {journal} {\bibinfo  {journal} {Nature Physics}\ }\textbf {\bibinfo {volume} {14}},\ \bibinfo {pages} {936--941} (\bibinfo {year} {2018})}\BibitemShut {NoStop}%
\bibitem [{\citenamefont {Halder}\ \emph {et~al.}(2018)\citenamefont {Halder}, \citenamefont {Chacon}, \citenamefont {Bauer}, \citenamefont {Simeth}, \citenamefont {M{\"u}hlbauer}, \citenamefont {Berger}, \citenamefont {Heinen}, \citenamefont {Garst}, \citenamefont {Rosch},\ and\ \citenamefont {Pfleiderer}}]{halder2018thermodynamic}%
  \BibitemOpen
  \bibfield  {author} {\bibinfo {author} {\bibfnamefont {M.}~\bibnamefont {Halder}}, \bibinfo {author} {\bibfnamefont {A.}~\bibnamefont {Chacon}}, \bibinfo {author} {\bibfnamefont {A.}~\bibnamefont {Bauer}}, \bibinfo {author} {\bibfnamefont {W.}~\bibnamefont {Simeth}}, \bibinfo {author} {\bibfnamefont {S.}~\bibnamefont {M{\"u}hlbauer}}, \bibinfo {author} {\bibfnamefont {H.}~\bibnamefont {Berger}}, \bibinfo {author} {\bibfnamefont {L.}~\bibnamefont {Heinen}}, \bibinfo {author} {\bibfnamefont {M.}~\bibnamefont {Garst}}, \bibinfo {author} {\bibfnamefont {A.}~\bibnamefont {Rosch}},\ and\ \bibinfo {author} {\bibfnamefont {C.}~\bibnamefont {Pfleiderer}},\ }\bibfield  {title} {\enquote {\bibinfo {title} {Thermodynamic evidence of a second skyrmion lattice phase and tilted conical phase in {Cu$_2$OSeO$_3$}},}\ }\href@noop {} {\bibfield  {journal} {\bibinfo  {journal} {Physical Review B}\ }\textbf {\bibinfo {volume} {98}},\ \bibinfo {pages} {144429} (\bibinfo {year} {2018})}\BibitemShut {NoStop}%
\bibitem [{\citenamefont {Bannenberg}\ \emph {et~al.}(2019)\citenamefont {Bannenberg}, \citenamefont {Wilhelm}, \citenamefont {Cubitt}, \citenamefont {Labh}, \citenamefont {Schmidt}, \citenamefont {Leli{\`e}vre-Berna}, \citenamefont {Pappas}, \citenamefont {Mostovoy},\ and\ \citenamefont {Leonov}}]{bannenberg2019multiple}%
  \BibitemOpen
  \bibfield  {author} {\bibinfo {author} {\bibfnamefont {L.~J.}\ \bibnamefont {Bannenberg}}, \bibinfo {author} {\bibfnamefont {H.}~\bibnamefont {Wilhelm}}, \bibinfo {author} {\bibfnamefont {R.}~\bibnamefont {Cubitt}}, \bibinfo {author} {\bibfnamefont {A.}~\bibnamefont {Labh}}, \bibinfo {author} {\bibfnamefont {M.~P.}\ \bibnamefont {Schmidt}}, \bibinfo {author} {\bibfnamefont {E.}~\bibnamefont {Leli{\`e}vre-Berna}}, \bibinfo {author} {\bibfnamefont {C.}~\bibnamefont {Pappas}}, \bibinfo {author} {\bibfnamefont {M.}~\bibnamefont {Mostovoy}},\ and\ \bibinfo {author} {\bibfnamefont {A.~O.}\ \bibnamefont {Leonov}},\ }\bibfield  {title} {\enquote {\bibinfo {title} {Multiple low-temperature skyrmionic states in a bulk chiral magnet},}\ }\href@noop {} {\bibfield  {journal} {\bibinfo  {journal} {npj Quantum Materials}\ }\textbf {\bibinfo {volume} {4}},\ \bibinfo {pages} {11} (\bibinfo {year} {2019})}\BibitemShut {NoStop}%
\bibitem [{\citenamefont {Marchiori}\ \emph {et~al.}(2024)\citenamefont {Marchiori}, \citenamefont {Romagnoli}, \citenamefont {Schneider}, \citenamefont {Gross}, \citenamefont {Sahafi}, \citenamefont {Jordan}, \citenamefont {Budakian}, \citenamefont {Baral}, \citenamefont {Magrez}, \citenamefont {White} \emph {et~al.}}]{marchiori2024imaging}%
  \BibitemOpen
  \bibfield  {author} {\bibinfo {author} {\bibfnamefont {E.}~\bibnamefont {Marchiori}}, \bibinfo {author} {\bibfnamefont {G.}~\bibnamefont {Romagnoli}}, \bibinfo {author} {\bibfnamefont {L.}~\bibnamefont {Schneider}}, \bibinfo {author} {\bibfnamefont {B.}~\bibnamefont {Gross}}, \bibinfo {author} {\bibfnamefont {P.}~\bibnamefont {Sahafi}}, \bibinfo {author} {\bibfnamefont {A.}~\bibnamefont {Jordan}}, \bibinfo {author} {\bibfnamefont {R.}~\bibnamefont {Budakian}}, \bibinfo {author} {\bibfnamefont {P.~R.}\ \bibnamefont {Baral}}, \bibinfo {author} {\bibfnamefont {A.}~\bibnamefont {Magrez}}, \bibinfo {author} {\bibfnamefont {J.~S.}\ \bibnamefont {White}}, \emph {et~al.},\ }\bibfield  {title} {\enquote {\bibinfo {title} {Imaging magnetic spiral phases, skyrmion clusters, and skyrmion displacements at the surface of bulk {Cu$_2$OSeO$_3$}},}\ }\href@noop {} {\bibfield  {journal} {\bibinfo  {journal} {Communications Materials}\ }\textbf {\bibinfo {volume} {5}},\ \bibinfo {pages} {202} (\bibinfo {year}
  {2024})}\BibitemShut {NoStop}%
\bibitem [{\citenamefont {Mehboodi}\ \emph {et~al.}(2024)\citenamefont {Mehboodi}, \citenamefont {Ukleev}, \citenamefont {Luo}, \citenamefont {Abrudan}, \citenamefont {Radu}, \citenamefont {Back},\ and\ \citenamefont {Aqeel}}]{mehboodi2024observation}%
  \BibitemOpen
  \bibfield  {author} {\bibinfo {author} {\bibfnamefont {S.}~\bibnamefont {Mehboodi}}, \bibinfo {author} {\bibfnamefont {V.}~\bibnamefont {Ukleev}}, \bibinfo {author} {\bibfnamefont {C.}~\bibnamefont {Luo}}, \bibinfo {author} {\bibfnamefont {R.}~\bibnamefont {Abrudan}}, \bibinfo {author} {\bibfnamefont {F.}~\bibnamefont {Radu}}, \bibinfo {author} {\bibfnamefont {C.}~\bibnamefont {Back}},\ and\ \bibinfo {author} {\bibfnamefont {A.}~\bibnamefont {Aqeel}},\ }\bibfield  {title} {\enquote {\bibinfo {title} {Observation of distorted tilted conical phase at the surface of a bulk chiral magnet with resonant elastic x-ray scattering},}\ }\href@noop {} {\bibfield  {journal} {\bibinfo  {journal} {arXiv preprint arXiv:2412.15882}\ } (\bibinfo {year} {2024})}\BibitemShut {NoStop}%
\bibitem [{\citenamefont {Baral}\ \emph {et~al.}(2022)\citenamefont {Baral}, \citenamefont {Ukleev}, \citenamefont {LaGrange}, \citenamefont {Cubitt}, \citenamefont {Zivkovic}, \citenamefont {R{\o}nnow}, \citenamefont {White},\ and\ \citenamefont {Magrez}}]{baral2022tuning}%
  \BibitemOpen
  \bibfield  {author} {\bibinfo {author} {\bibfnamefont {P.~R.}\ \bibnamefont {Baral}}, \bibinfo {author} {\bibfnamefont {V.}~\bibnamefont {Ukleev}}, \bibinfo {author} {\bibfnamefont {T.}~\bibnamefont {LaGrange}}, \bibinfo {author} {\bibfnamefont {R.}~\bibnamefont {Cubitt}}, \bibinfo {author} {\bibfnamefont {I.}~\bibnamefont {Zivkovic}}, \bibinfo {author} {\bibfnamefont {H.~M.}\ \bibnamefont {R{\o}nnow}}, \bibinfo {author} {\bibfnamefont {J.~S.}\ \bibnamefont {White}},\ and\ \bibinfo {author} {\bibfnamefont {A.}~\bibnamefont {Magrez}},\ }\bibfield  {title} {\enquote {\bibinfo {title} {Tuning topological spin textures in size-tailored chiral magnet insulator particles},}\ }\href@noop {} {\bibfield  {journal} {\bibinfo  {journal} {The Journal of Physical Chemistry C}\ }\textbf {\bibinfo {volume} {126}},\ \bibinfo {pages} {11855--11866} (\bibinfo {year} {2022})}\BibitemShut {NoStop}%
\bibitem [{\citenamefont {Janson}\ \emph {et~al.}(2014)\citenamefont {Janson}, \citenamefont {Rousochatzakis}, \citenamefont {Tsirlin}, \citenamefont {Belesi}, \citenamefont {Leonov}, \citenamefont {R{\"o}{\ss}ler}, \citenamefont {Van Den~Brink},\ and\ \citenamefont {Rosner}}]{janson2014quantum}%
  \BibitemOpen
  \bibfield  {author} {\bibinfo {author} {\bibfnamefont {O.}~\bibnamefont {Janson}}, \bibinfo {author} {\bibfnamefont {I.}~\bibnamefont {Rousochatzakis}}, \bibinfo {author} {\bibfnamefont {A.~A.}\ \bibnamefont {Tsirlin}}, \bibinfo {author} {\bibfnamefont {M.}~\bibnamefont {Belesi}}, \bibinfo {author} {\bibfnamefont {A.~A.}\ \bibnamefont {Leonov}}, \bibinfo {author} {\bibfnamefont {U.~K.}\ \bibnamefont {R{\"o}{\ss}ler}}, \bibinfo {author} {\bibfnamefont {J.}~\bibnamefont {Van Den~Brink}},\ and\ \bibinfo {author} {\bibfnamefont {H.}~\bibnamefont {Rosner}},\ }\bibfield  {title} {\enquote {\bibinfo {title} {The quantum nature of skyrmions and half-skyrmions in {Cu$_2$OSeO$_3$}},}\ }\href@noop {} {\bibfield  {journal} {\bibinfo  {journal} {Nature Communications}\ }\textbf {\bibinfo {volume} {5}},\ \bibinfo {pages} {5376} (\bibinfo {year} {2014})}\BibitemShut {NoStop}%
\bibitem [{\citenamefont {Grytsiuk}\ \emph {et~al.}(2019)\citenamefont {Grytsiuk}, \citenamefont {Hoffmann}, \citenamefont {Hanke}, \citenamefont {Mavropoulos}, \citenamefont {Mokrousov}, \citenamefont {Bihlmayer},\ and\ \citenamefont {Bl{\"u}gel}}]{grytsiuk2019ab}%
  \BibitemOpen
  \bibfield  {author} {\bibinfo {author} {\bibfnamefont {S.}~\bibnamefont {Grytsiuk}}, \bibinfo {author} {\bibfnamefont {M.}~\bibnamefont {Hoffmann}}, \bibinfo {author} {\bibfnamefont {J.-P.}\ \bibnamefont {Hanke}}, \bibinfo {author} {\bibfnamefont {P.}~\bibnamefont {Mavropoulos}}, \bibinfo {author} {\bibfnamefont {Y.}~\bibnamefont {Mokrousov}}, \bibinfo {author} {\bibfnamefont {G.}~\bibnamefont {Bihlmayer}},\ and\ \bibinfo {author} {\bibfnamefont {S.}~\bibnamefont {Bl{\"u}gel}},\ }\bibfield  {title} {\enquote {\bibinfo {title} {Ab initio analysis of magnetic properties of the prototype {B20 chiral magnet FeGe}},}\ }\href@noop {} {\bibfield  {journal} {\bibinfo  {journal} {Physical Review B}\ }\textbf {\bibinfo {volume} {100}},\ \bibinfo {pages} {214406} (\bibinfo {year} {2019})}\BibitemShut {NoStop}%
\bibitem [{\citenamefont {Grigoriev}\ \emph {et~al.}(2015)\citenamefont {Grigoriev}, \citenamefont {Sukhanov}, \citenamefont {Altynbaev}, \citenamefont {Siegfried}, \citenamefont {Heinemann}, \citenamefont {Kizhe},\ and\ \citenamefont {Maleyev}}]{grigoriev2015spin}%
  \BibitemOpen
  \bibfield  {author} {\bibinfo {author} {\bibfnamefont {S.}~\bibnamefont {Grigoriev}}, \bibinfo {author} {\bibfnamefont {A.}~\bibnamefont {Sukhanov}}, \bibinfo {author} {\bibfnamefont {E.}~\bibnamefont {Altynbaev}}, \bibinfo {author} {\bibfnamefont {S.-A.}\ \bibnamefont {Siegfried}}, \bibinfo {author} {\bibfnamefont {A.}~\bibnamefont {Heinemann}}, \bibinfo {author} {\bibfnamefont {P.}~\bibnamefont {Kizhe}},\ and\ \bibinfo {author} {\bibfnamefont {S.}~\bibnamefont {Maleyev}},\ }\bibfield  {title} {\enquote {\bibinfo {title} {Spin waves in full-polarized state of {Dzyaloshinskii-Moriya helimagnets: Small-angle neutron scattering study}},}\ }\href@noop {} {\bibfield  {journal} {\bibinfo  {journal} {Physical Review B}\ }\textbf {\bibinfo {volume} {92}},\ \bibinfo {pages} {220415} (\bibinfo {year} {2015})}\BibitemShut {NoStop}%
\bibitem [{\citenamefont {Seki}\ \emph {et~al.}(2012)\citenamefont {Seki}, \citenamefont {Yu}, \citenamefont {Ishiwata},\ and\ \citenamefont {Tokura}}]{seki2012observation}%
  \BibitemOpen
  \bibfield  {author} {\bibinfo {author} {\bibfnamefont {S.}~\bibnamefont {Seki}}, \bibinfo {author} {\bibfnamefont {X.}~\bibnamefont {Yu}}, \bibinfo {author} {\bibfnamefont {S.}~\bibnamefont {Ishiwata}},\ and\ \bibinfo {author} {\bibfnamefont {Y.}~\bibnamefont {Tokura}},\ }\bibfield  {title} {\enquote {\bibinfo {title} {Observation of skyrmions in a multiferroic material},}\ }\href@noop {} {\bibfield  {journal} {\bibinfo  {journal} {Science}\ }\textbf {\bibinfo {volume} {336}},\ \bibinfo {pages} {198--201} (\bibinfo {year} {2012})}\BibitemShut {NoStop}%
\bibitem [{\citenamefont {Dewhurst}\ \emph {et~al.}(2016)\citenamefont {Dewhurst}, \citenamefont {Grillo}, \citenamefont {Honecker}, \citenamefont {Bonnaud}, \citenamefont {Jacques}, \citenamefont {Amrouni}, \citenamefont {Perillo-Marcone}, \citenamefont {Manzin},\ and\ \citenamefont {Cubitt}}]{dewhurst2016small}%
  \BibitemOpen
  \bibfield  {author} {\bibinfo {author} {\bibfnamefont {C.}~\bibnamefont {Dewhurst}}, \bibinfo {author} {\bibfnamefont {I.}~\bibnamefont {Grillo}}, \bibinfo {author} {\bibfnamefont {D.}~\bibnamefont {Honecker}}, \bibinfo {author} {\bibfnamefont {M.}~\bibnamefont {Bonnaud}}, \bibinfo {author} {\bibfnamefont {M.}~\bibnamefont {Jacques}}, \bibinfo {author} {\bibfnamefont {C.}~\bibnamefont {Amrouni}}, \bibinfo {author} {\bibfnamefont {A.}~\bibnamefont {Perillo-Marcone}}, \bibinfo {author} {\bibfnamefont {G.}~\bibnamefont {Manzin}},\ and\ \bibinfo {author} {\bibfnamefont {R.}~\bibnamefont {Cubitt}},\ }\bibfield  {title} {\enquote {\bibinfo {title} {{The small-angle neutron scattering instrument D33 at the Institut Laue-Langevin}},}\ }\href@noop {} {\bibfield  {journal} {\bibinfo  {journal} {Journal of Applied Crystallography}\ }\textbf {\bibinfo {volume} {49}},\ \bibinfo {pages} {1--14} (\bibinfo {year} {2016})}\BibitemShut {NoStop}%
\bibitem [{\citenamefont {Dewhurst}(2023)}]{dewhurst2023graphical}%
  \BibitemOpen
  \bibfield  {author} {\bibinfo {author} {\bibfnamefont {C.}~\bibnamefont {Dewhurst}},\ }\bibfield  {title} {\enquote {\bibinfo {title} {Graphical reduction and analysis small-angle neutron scattering program: {GRASP}},}\ }\href@noop {} {\bibfield  {journal} {\bibinfo  {journal} {Journal of Applied Crystallography}\ }\textbf {\bibinfo {volume} {56}} (\bibinfo {year} {2023})}\BibitemShut {NoStop}%
\bibitem [{\citenamefont {Kataoka}(1987)}]{kataoka1987}%
  \BibitemOpen
  \bibfield  {author} {\bibinfo {author} {\bibfnamefont {M.}~\bibnamefont {Kataoka}},\ }\bibfield  {title} {\enquote {\bibinfo {title} {Spin waves in systems with long period helical spin density waves due to the antisymmetric and symmetric exchange interactions},}\ }\href@noop {} {\bibfield  {journal} {\bibinfo  {journal} {Journal of the Physical Society of Japan}\ }\textbf {\bibinfo {volume} {56}},\ \bibinfo {pages} {3635--3647} (\bibinfo {year} {1987})}\BibitemShut {NoStop}%
\bibitem [{\citenamefont {Baral}\ \emph {et~al.}(2023{\natexlab{a}})\citenamefont {Baral}, \citenamefont {Utesov}, \citenamefont {Luo}, \citenamefont {Radu}, \citenamefont {Magrez}, \citenamefont {White},\ and\ \citenamefont {Ukleev}}]{baral202023direct}%
  \BibitemOpen
  \bibfield  {author} {\bibinfo {author} {\bibfnamefont {P.~R.}\ \bibnamefont {Baral}}, \bibinfo {author} {\bibfnamefont {O.~I.}\ \bibnamefont {Utesov}}, \bibinfo {author} {\bibfnamefont {C.}~\bibnamefont {Luo}}, \bibinfo {author} {\bibfnamefont {F.}~\bibnamefont {Radu}}, \bibinfo {author} {\bibfnamefont {A.}~\bibnamefont {Magrez}}, \bibinfo {author} {\bibfnamefont {J.~S.}\ \bibnamefont {White}},\ and\ \bibinfo {author} {\bibfnamefont {V.}~\bibnamefont {Ukleev}},\ }\bibfield  {title} {\enquote {\bibinfo {title} {Direct observation of exchange anisotropy in the helimagnetic insulator {${\mathrm{Cu}}_{2}{\mathrm{OSeO}}_{3}$}},}\ }\href@noop {} {\bibfield  {journal} {\bibinfo  {journal} {Physical Review Research}\ }\textbf {\bibinfo {volume} {5}},\ \bibinfo {pages} {L032019} (\bibinfo {year} {2023}{\natexlab{a}})}\BibitemShut {NoStop}%
\bibitem [{\citenamefont {Adams}\ \emph {et~al.}(2012)\citenamefont {Adams}, \citenamefont {Chacon}, \citenamefont {Wagner}, \citenamefont {Bauer}, \citenamefont {Brandl}, \citenamefont {Pedersen}, \citenamefont {Berger}, \citenamefont {Lemmens},\ and\ \citenamefont {Pfleiderer}}]{adams2012long}%
  \BibitemOpen
  \bibfield  {author} {\bibinfo {author} {\bibfnamefont {T.}~\bibnamefont {Adams}}, \bibinfo {author} {\bibfnamefont {A.}~\bibnamefont {Chacon}}, \bibinfo {author} {\bibfnamefont {M.}~\bibnamefont {Wagner}}, \bibinfo {author} {\bibfnamefont {A.}~\bibnamefont {Bauer}}, \bibinfo {author} {\bibfnamefont {G.}~\bibnamefont {Brandl}}, \bibinfo {author} {\bibfnamefont {B.}~\bibnamefont {Pedersen}}, \bibinfo {author} {\bibfnamefont {H.}~\bibnamefont {Berger}}, \bibinfo {author} {\bibfnamefont {P.}~\bibnamefont {Lemmens}},\ and\ \bibinfo {author} {\bibfnamefont {C.}~\bibnamefont {Pfleiderer}},\ }\bibfield  {title} {\enquote {\bibinfo {title} {Long-wavelength helimagnetic order and skyrmion lattice phase in {Cu$_2$OSeO$_3$}},}\ }\href@noop {} {\bibfield  {journal} {\bibinfo  {journal} {Physical Review Letters}\ }\textbf {\bibinfo {volume} {108}},\ \bibinfo {pages} {237204} (\bibinfo {year} {2012})}\BibitemShut {NoStop}%
\bibitem [{\citenamefont {Ukleev}\ \emph {et~al.}(2022)\citenamefont {Ukleev}, \citenamefont {Pschenichnyi}, \citenamefont {Utesov}, \citenamefont {Karube}, \citenamefont {M{\"u}hlbauer}, \citenamefont {Cubitt}, \citenamefont {Tokura}, \citenamefont {Taguchi}, \citenamefont {White},\ and\ \citenamefont {Grigoriev}}]{ukleev2022spin}%
  \BibitemOpen
  \bibfield  {author} {\bibinfo {author} {\bibfnamefont {V.}~\bibnamefont {Ukleev}}, \bibinfo {author} {\bibfnamefont {K.}~\bibnamefont {Pschenichnyi}}, \bibinfo {author} {\bibfnamefont {O.}~\bibnamefont {Utesov}}, \bibinfo {author} {\bibfnamefont {K.}~\bibnamefont {Karube}}, \bibinfo {author} {\bibfnamefont {S.}~\bibnamefont {M{\"u}hlbauer}}, \bibinfo {author} {\bibfnamefont {R.}~\bibnamefont {Cubitt}}, \bibinfo {author} {\bibfnamefont {Y.}~\bibnamefont {Tokura}}, \bibinfo {author} {\bibfnamefont {Y.}~\bibnamefont {Taguchi}}, \bibinfo {author} {\bibfnamefont {J.}~\bibnamefont {White}},\ and\ \bibinfo {author} {\bibfnamefont {S.}~\bibnamefont {Grigoriev}},\ }\bibfield  {title} {\enquote {\bibinfo {title} {Spin wave stiffness and damping in a frustrated chiral helimagnet {Co$_8$Zn$_8$Mn$_4$} as measured by small-angle neutron scattering},}\ }\href@noop {} {\bibfield  {journal} {\bibinfo  {journal} {Physical Review Research}\ }\textbf {\bibinfo {volume} {4}},\ \bibinfo {pages} {023239} (\bibinfo {year}
  {2022})}\BibitemShut {NoStop}%
\bibitem [{\citenamefont {Grigoriev}\ \emph {et~al.}(2019{\natexlab{a}})\citenamefont {Grigoriev}, \citenamefont {Pschenichnyi}, \citenamefont {Altynbaev}, \citenamefont {Heinemann},\ and\ \citenamefont {Magrez}}]{grigoriev2019spin}%
  \BibitemOpen
  \bibfield  {author} {\bibinfo {author} {\bibfnamefont {S.}~\bibnamefont {Grigoriev}}, \bibinfo {author} {\bibfnamefont {K.}~\bibnamefont {Pschenichnyi}}, \bibinfo {author} {\bibfnamefont {E.}~\bibnamefont {Altynbaev}}, \bibinfo {author} {\bibfnamefont {A.}~\bibnamefont {Heinemann}},\ and\ \bibinfo {author} {\bibfnamefont {A.}~\bibnamefont {Magrez}},\ }\bibfield  {title} {\enquote {\bibinfo {title} {Spin-wave stiffness in the {Dzyaloshinskii-Moriya} helimagnet with ferrimagnetic ordering {Cu$_2$OSeO$_3$}},}\ }\href@noop {} {\bibfield  {journal} {\bibinfo  {journal} {Physical Review B}\ }\textbf {\bibinfo {volume} {99}},\ \bibinfo {pages} {054427} (\bibinfo {year} {2019}{\natexlab{a}})}\BibitemShut {NoStop}%
\bibitem [{\citenamefont {Grigoriev}\ \emph {et~al.}(2018)\citenamefont {Grigoriev}, \citenamefont {Altynbaev}, \citenamefont {Siegfried}, \citenamefont {Pschenichnyi}, \citenamefont {Menzel}, \citenamefont {Heinemann},\ and\ \citenamefont {Chaboussant}}]{grigoriev2018spin}%
  \BibitemOpen
  \bibfield  {author} {\bibinfo {author} {\bibfnamefont {S.}~\bibnamefont {Grigoriev}}, \bibinfo {author} {\bibfnamefont {E.}~\bibnamefont {Altynbaev}}, \bibinfo {author} {\bibfnamefont {S.-A.}\ \bibnamefont {Siegfried}}, \bibinfo {author} {\bibfnamefont {K.}~\bibnamefont {Pschenichnyi}}, \bibinfo {author} {\bibfnamefont {D.}~\bibnamefont {Menzel}}, \bibinfo {author} {\bibfnamefont {A.}~\bibnamefont {Heinemann}},\ and\ \bibinfo {author} {\bibfnamefont {G.}~\bibnamefont {Chaboussant}},\ }\bibfield  {title} {\enquote {\bibinfo {title} {Spin-wave stiffness in the dzyaloshinskii-moriya helimagnets {Mn$_{1-x}$Fe$_x$Si}},}\ }\href@noop {} {\bibfield  {journal} {\bibinfo  {journal} {Physical Review B}\ }\textbf {\bibinfo {volume} {97}},\ \bibinfo {pages} {024409} (\bibinfo {year} {2018})}\BibitemShut {NoStop}%
\bibitem [{\citenamefont {Siegfried}\ \emph {et~al.}(2017)\citenamefont {Siegfried}, \citenamefont {Sukhanov}, \citenamefont {Altynbaev}, \citenamefont {Honecker}, \citenamefont {Heinemann}, \citenamefont {Tsvyashchenko},\ and\ \citenamefont {Grigoriev}}]{siegfried2017spin}%
  \BibitemOpen
  \bibfield  {author} {\bibinfo {author} {\bibfnamefont {S.-A.}\ \bibnamefont {Siegfried}}, \bibinfo {author} {\bibfnamefont {A.}~\bibnamefont {Sukhanov}}, \bibinfo {author} {\bibfnamefont {E.}~\bibnamefont {Altynbaev}}, \bibinfo {author} {\bibfnamefont {D.}~\bibnamefont {Honecker}}, \bibinfo {author} {\bibfnamefont {A.}~\bibnamefont {Heinemann}}, \bibinfo {author} {\bibfnamefont {A.}~\bibnamefont {Tsvyashchenko}},\ and\ \bibinfo {author} {\bibfnamefont {S.}~\bibnamefont {Grigoriev}},\ }\bibfield  {title} {\enquote {\bibinfo {title} {Spin-wave dynamics in the helimagnet {FeGe} studied by small-angle neutron scattering},}\ }\href@noop {} {\bibfield  {journal} {\bibinfo  {journal} {Physical Review B}\ }\textbf {\bibinfo {volume} {95}},\ \bibinfo {pages} {134415} (\bibinfo {year} {2017})}\BibitemShut {NoStop}%
\bibitem [{\citenamefont {Grigoriev}\ \emph {et~al.}(2019{\natexlab{b}})\citenamefont {Grigoriev}, \citenamefont {Pschenichnyi}, \citenamefont {Altynbaev}, \citenamefont {Siegfried}, \citenamefont {Heinemann}, \citenamefont {Honnecker},\ and\ \citenamefont {Menzel}}]{grigoriev2019spin2}%
  \BibitemOpen
  \bibfield  {author} {\bibinfo {author} {\bibfnamefont {S.}~\bibnamefont {Grigoriev}}, \bibinfo {author} {\bibfnamefont {K.}~\bibnamefont {Pschenichnyi}}, \bibinfo {author} {\bibfnamefont {E.}~\bibnamefont {Altynbaev}}, \bibinfo {author} {\bibfnamefont {S.-A.}\ \bibnamefont {Siegfried}}, \bibinfo {author} {\bibfnamefont {A.}~\bibnamefont {Heinemann}}, \bibinfo {author} {\bibfnamefont {D.}~\bibnamefont {Honnecker}},\ and\ \bibinfo {author} {\bibfnamefont {D.}~\bibnamefont {Menzel}},\ }\bibfield  {title} {\enquote {\bibinfo {title} {Spin-wave stiffness of the {Dzyaloshinskii-Moriya} helimagnet compounds {Fe$_{1-x}$Co$_x$Si} studied by small-angle neutron scattering},}\ }\href@noop {} {\bibfield  {journal} {\bibinfo  {journal} {Physical Review B}\ }\textbf {\bibinfo {volume} {100}},\ \bibinfo {pages} {094409} (\bibinfo {year} {2019}{\natexlab{b}})}\BibitemShut {NoStop}%
\bibitem [{\citenamefont {Okorokov}\ \emph {et~al.}(1986)\citenamefont {Okorokov}, \citenamefont {Runov}, \citenamefont {Toperverg}, \citenamefont {Tret'Yakov}, \citenamefont {Mal'Tsev}, \citenamefont {Puzeǐ},\ and\ \citenamefont {Mikhaǐlova}}]{okorokov1986study}%
  \BibitemOpen
  \bibfield  {author} {\bibinfo {author} {\bibfnamefont {A.}~\bibnamefont {Okorokov}}, \bibinfo {author} {\bibfnamefont {V.}~\bibnamefont {Runov}}, \bibinfo {author} {\bibfnamefont {B.}~\bibnamefont {Toperverg}}, \bibinfo {author} {\bibfnamefont {A.}~\bibnamefont {Tret'Yakov}}, \bibinfo {author} {\bibfnamefont {E.}~\bibnamefont {Mal'Tsev}}, \bibinfo {author} {\bibfnamefont {I.}~\bibnamefont {Puzeǐ}},\ and\ \bibinfo {author} {\bibfnamefont {V.}~\bibnamefont {Mikhaǐlova}},\ }\bibfield  {title} {\enquote {\bibinfo {title} {Study of spin waves in amorphous magnetic materials by polarized-neutron scattering},}\ }\href@noop {} {\bibfield  {journal} {\bibinfo  {journal} {Soviet Journal of Experimental and Theoretical Physics Letters}\ }\textbf {\bibinfo {volume} {43}},\ \bibinfo {pages} {503} (\bibinfo {year} {1986})}\BibitemShut {NoStop}%
\bibitem [{\citenamefont {Deriglazov}\ \emph {et~al.}(1992)\citenamefont {Deriglazov}, \citenamefont {Okorokov}, \citenamefont {Runov}, \citenamefont {Toperverg}, \citenamefont {Kampmann}, \citenamefont {Eckerlebe}, \citenamefont {Schmidt},\ and\ \citenamefont {L{\"o}bner}}]{deriglazov1992study}%
  \BibitemOpen
  \bibfield  {author} {\bibinfo {author} {\bibfnamefont {V.}~\bibnamefont {Deriglazov}}, \bibinfo {author} {\bibfnamefont {A.}~\bibnamefont {Okorokov}}, \bibinfo {author} {\bibfnamefont {V.}~\bibnamefont {Runov}}, \bibinfo {author} {\bibfnamefont {B.}~\bibnamefont {Toperverg}}, \bibinfo {author} {\bibfnamefont {R.}~\bibnamefont {Kampmann}}, \bibinfo {author} {\bibfnamefont {H.}~\bibnamefont {Eckerlebe}}, \bibinfo {author} {\bibfnamefont {W.}~\bibnamefont {Schmidt}},\ and\ \bibinfo {author} {\bibfnamefont {W.}~\bibnamefont {L{\"o}bner}},\ }\bibfield  {title} {\enquote {\bibinfo {title} {Study of spin waves in amorphous ferromagnet {Fe$_{50}$Ni$_{22}$Cr$_{10}$P$_{18}$} by small angle polarized neutron scattering},}\ }\href@noop {} {\bibfield  {journal} {\bibinfo  {journal} {Physica B: Condensed Matter}\ }\textbf {\bibinfo {volume} {180}},\ \bibinfo {pages} {262--264} (\bibinfo {year} {1992})}\BibitemShut {NoStop}%
\bibitem [{\citenamefont {Luo}\ \emph {et~al.}(2020)\citenamefont {Luo}, \citenamefont {Marcus}, \citenamefont {Trump}, \citenamefont {Kindervater}, \citenamefont {Stone}, \citenamefont {Rodriguez-Rivera}, \citenamefont {Qiu}, \citenamefont {McQueen}, \citenamefont {Tchernyshyov},\ and\ \citenamefont {Broholm}}]{luo2020low}%
  \BibitemOpen
  \bibfield  {author} {\bibinfo {author} {\bibfnamefont {Y.}~\bibnamefont {Luo}}, \bibinfo {author} {\bibfnamefont {G.}~\bibnamefont {Marcus}}, \bibinfo {author} {\bibfnamefont {B.}~\bibnamefont {Trump}}, \bibinfo {author} {\bibfnamefont {J.}~\bibnamefont {Kindervater}}, \bibinfo {author} {\bibfnamefont {M.}~\bibnamefont {Stone}}, \bibinfo {author} {\bibfnamefont {J.}~\bibnamefont {Rodriguez-Rivera}}, \bibinfo {author} {\bibfnamefont {Y.}~\bibnamefont {Qiu}}, \bibinfo {author} {\bibfnamefont {T.}~\bibnamefont {McQueen}}, \bibinfo {author} {\bibfnamefont {O.}~\bibnamefont {Tchernyshyov}},\ and\ \bibinfo {author} {\bibfnamefont {C.}~\bibnamefont {Broholm}},\ }\bibfield  {title} {\enquote {\bibinfo {title} {Low-energy magnons in the chiral ferrimagnet {Cu$_2$OSeO$_3$}: A coarse-grained approach},}\ }\href@noop {} {\bibfield  {journal} {\bibinfo  {journal} {Physical Review B}\ }\textbf {\bibinfo {volume} {101}},\ \bibinfo {pages} {144411} (\bibinfo {year} {2020})}\BibitemShut {NoStop}%
\bibitem [{\citenamefont {Portnichenko}\ \emph {et~al.}(2016)\citenamefont {Portnichenko}, \citenamefont {Romh{\'a}nyi}, \citenamefont {Onykiienko}, \citenamefont {Henschel}, \citenamefont {Schmidt}, \citenamefont {Cameron}, \citenamefont {Surmach}, \citenamefont {Lim}, \citenamefont {Park}, \citenamefont {Schneidewind} \emph {et~al.}}]{portnichenko2016magnon}%
  \BibitemOpen
  \bibfield  {author} {\bibinfo {author} {\bibfnamefont {P.}~\bibnamefont {Portnichenko}}, \bibinfo {author} {\bibfnamefont {J.}~\bibnamefont {Romh{\'a}nyi}}, \bibinfo {author} {\bibfnamefont {Y.}~\bibnamefont {Onykiienko}}, \bibinfo {author} {\bibfnamefont {A.}~\bibnamefont {Henschel}}, \bibinfo {author} {\bibfnamefont {M.}~\bibnamefont {Schmidt}}, \bibinfo {author} {\bibfnamefont {A.}~\bibnamefont {Cameron}}, \bibinfo {author} {\bibfnamefont {M.}~\bibnamefont {Surmach}}, \bibinfo {author} {\bibfnamefont {J.}~\bibnamefont {Lim}}, \bibinfo {author} {\bibfnamefont {J.}~\bibnamefont {Park}}, \bibinfo {author} {\bibfnamefont {A.}~\bibnamefont {Schneidewind}}, \emph {et~al.},\ }\bibfield  {title} {\enquote {\bibinfo {title} {Magnon spectrum of the helimagnetic insulator {Cu$_2$OSeO$_3$}},}\ }\href@noop {} {\bibfield  {journal} {\bibinfo  {journal} {Nature Communications}\ }\textbf {\bibinfo {volume} {7}},\ \bibinfo {pages} {10725} (\bibinfo {year} {2016})}\BibitemShut {NoStop}%
\bibitem [{\citenamefont {Tucker}\ \emph {et~al.}(2016)\citenamefont {Tucker}, \citenamefont {White}, \citenamefont {Romh{\'a}nyi}, \citenamefont {Szaller}, \citenamefont {K{\'e}zsm{\'a}rki}, \citenamefont {Roessli}, \citenamefont {Stuhr}, \citenamefont {Magrez}, \citenamefont {Groitl}, \citenamefont {Babkevich} \emph {et~al.}}]{tucker2016spin}%
  \BibitemOpen
  \bibfield  {author} {\bibinfo {author} {\bibfnamefont {G.}~\bibnamefont {Tucker}}, \bibinfo {author} {\bibfnamefont {J.}~\bibnamefont {White}}, \bibinfo {author} {\bibfnamefont {J.}~\bibnamefont {Romh{\'a}nyi}}, \bibinfo {author} {\bibfnamefont {D.}~\bibnamefont {Szaller}}, \bibinfo {author} {\bibfnamefont {I.}~\bibnamefont {K{\'e}zsm{\'a}rki}}, \bibinfo {author} {\bibfnamefont {B.}~\bibnamefont {Roessli}}, \bibinfo {author} {\bibfnamefont {U.}~\bibnamefont {Stuhr}}, \bibinfo {author} {\bibfnamefont {A.}~\bibnamefont {Magrez}}, \bibinfo {author} {\bibfnamefont {F.}~\bibnamefont {Groitl}}, \bibinfo {author} {\bibfnamefont {P.}~\bibnamefont {Babkevich}}, \emph {et~al.},\ }\bibfield  {title} {\enquote {\bibinfo {title} {Spin excitations in the skyrmion host {Cu$_2$OSeO$_3$}},}\ }\href@noop {} {\bibfield  {journal} {\bibinfo  {journal} {Physical Review B}\ }\textbf {\bibinfo {volume} {93}},\ \bibinfo {pages} {054401} (\bibinfo {year} {2016})}\BibitemShut {NoStop}%
\bibitem [{\citenamefont {Baral}\ \emph {et~al.}(2023{\natexlab{b}})\citenamefont {Baral}, \citenamefont {Cubitt}, \citenamefont {Steinke}, \citenamefont {Ukleev},\ and\ \citenamefont {White}}]{illdata}%
  \BibitemOpen
  \bibfield  {author} {\bibinfo {author} {\bibfnamefont {P.}~\bibnamefont {Baral}}, \bibinfo {author} {\bibfnamefont {R.}~\bibnamefont {Cubitt}}, \bibinfo {author} {\bibfnamefont {{\relax N.J}.}~\bibnamefont {Steinke}}, \bibinfo {author} {\bibfnamefont {V.}~\bibnamefont {Ukleev}},\ and\ \bibinfo {author} {\bibfnamefont {{\relax J.S}.}~\bibnamefont {White}},\ }\href@noop {} {\enquote {\bibinfo {title} {Determining the magnetic structure of a novel cubic chiral insulator using polarized and unpolarized {SANS}},}\ }\bibinfo {howpublished} {Institut Laue-Langevin (ILL) \url{https://doi.ill.fr/10.5291/ILL-DATA.5-41-1226}} (\bibinfo {year} {2023}{\natexlab{b}})\BibitemShut {NoStop}%
\end{thebibliography}%

\end{document}